\documentclass[a4paper]{article}

\usepackage{graphicx}
\usepackage{color}
\usepackage{url}
\usepackage{wrapfig}

\usepackage{fancyhdr}
\fancypagestyle{firststyle}
{ 
   \fancyhf{}%
   \fancyfoot[C]{\footnotesize Distribution Statement A. Approved for Public Release. Distribution Unlimited.}
}

\newcommand{\omitit}[1]{}

\title{A Comparison of State-of-the-Art Techniques for Generating Adversarial Malware Binaries}
\author{Prithviraj Dasgupta and Zachariah Osman{\footnote{P. Dasgupta is with the Distributed Intelligent Systems Section, Information Technology Division at the U.S. Naval Research Laboratory, Washington, D.C. Z. Osman is an undergraduate student in the computer science department at George Mason University. He worked as an NREIP student intern during summer 2021 at the Naval Research Laboratory, Washington, D. C.}}\\[0.1in]
U. S. Naval Research Laboratory, Washington, D. C.\\[0.1in]
Contact author email: \{raj.dasgupta@nrl.navy.mil\}}

\begin{document}
\maketitle

\thispagestyle{firststyle}

\begin{center}
{\bf Abstract}
\end{center}
We consider the problem of generating adversarial malware by a cyber-attacker where the attacker's task is to strategically  modify certain bytes within existing binary malware files, so that the modified files are able to evade a malware detector such as machine learning-based malware classifier. We have evaluated three recent adversarial malware generation techniques using binary malware samples drawn from a single, publicly available malware data set and compared their performances for evading a machine-learning based malware classifier called MalConv. Our results show that among the compared techniques, the most effective technique is the one that strategically modifies bytes in a binary's header. We conclude by discussing the lessons learned and future research directions on the topic of adversarial malware generation.

\section{Introduction}
\label{sec_intro}
Malicious software or malware have perennially been an Achilles heel for computer systems. Cyber security researchers have regularly devised sophisticated hardware and software measures to counter malware, but malware creators have relentlessly discovered and exploited newer vulnerabilities to compromise computer systems. In $2019$, nearly $50$ million computer viruses were detected~\cite{malwarebytes2020}, including a $13\%$ increase in malware threat activity for businesses. Conventional approaches to safeguard computer systems against malware used static analysis of binary files that use signature-based rules designed by humans to identify and flag suspicious behavior of an executable file. Static malware analysis techniques have been widely successful and are currently used in several anti-virus (AV) software tools. However, they require the malware signature rules to be continuously updated and are less resilient to previously unseen exploitation strategies. Dynamic or run-time behavior analysis techniques such as sandboxing for binary files are more successful than static analysis techniques, but they require considerable overhead in terms of run-time and computational resources to prevent malware from harming the computer system. Consequently, dynamic malware analysis techniques are not ideal for use in time- or resource-critical settings such as browser plugins, mobile devices or thin clients. Cleverly designed malware are also known to be able to detect if they are being executed inside a sandbox and suppress their malicious behavior. 

Over the past decade, researchers have started to investigate artificial intelligence (AI) and machine learning (ML) based techniques such as classifiers as a means to quickly discern and detect complex signature patterns that might indicate malicious behavior in binary files. Simultaneously, malware creators have also leveraged AI and ML techniques to intelligently craft malware that can evade ML classifiers. To defeat such AI-enabled, evasive malware, it makes sense to explore and understand strategies for automatically creating malware so that appropriate countermeasures could be developed against them. In this report, we analyze recent techniques from literature for adversarial malware generation that strategically modify bytes inside a malware to create an evasive version of the malware while retaining its malicious functionality. Recently proposed byte-level adversarial malware generation techniques have been evaluated in literature with different malware data sets and with different malware detection tools like commercial AV software and AI-based classifiers. As a result, it is not straightforward to compare their effectiveness and analyze their strengths and weaknesses in a systematic manner. Here, we attempt to address this limitation by evaluating three recent adversarial malware generation techniques using binary file samples drawn from a single, publicly available malware data set and compare their performances for performing evasion attacks on the same malware classifier. Our results show that among the compared techniques, the most effective technique across all performance measures is the one that strategically modifies bytes in the binary's header. Based on our research results and lessons we identify directions for future research that would help to build effective defenses towards making computer systems more robust against intelligently crafted malware attacks.

\section{Related Work}
\label{sec_relwork}
\subsection{Adversarial Malware Generation and Evading ML Classifiers}
Generating binary files that can evade machine learning-based classifiers has been active area of research for over a decade. Recently, researchers have proposed machine learning techniques that an adversary could use to strategically craft malware instances that can evade a malware classifier. Initial work on machine learning-based malware generation mainly focused on extracting and modifying features of the binary. Examples of binary file features include {\tt DebugSize, DebugRVA, ImageVersion, OperatingSystemVersion, LinkerVersion, DllCharacteristics, ExportFunctionsCount,} etc.~\cite{raman2012selecting}. A limitation of making modifications in the latent feature space of a binary is that, post-modification, it is difficult to reverse engineer the binary corresponding to the modified features. Recently, to address this limitation, researchers have started investigating techniques that directly modify the binary at the byte level to generate a modified binary that can evade an AV tool or machine learning-based malware classifier. We review two main directions of research for adversarial malware generation - techniques that modify the binary in the feature space using AI-based methods, and, techniques that modify the binary at the byte level, as described below:

{\bf Feature Space Modifications using Reinforcement Learning.} Features of binary files have been widely-used as a means in AV tools of classifying their maliciousness because the features the malware's signature. Consequently, strategically modifying features of a malware binary can modify their signature without altering their malicious functionality. Feature modification techniques first transform or embed the bytes of the malware binary into a latent feature space and then modify those features~\cite{hu2017generating}. However, it must be noted that it is non-trivial to reverse transform the modified features into their byte-level equivalents to get the modified binary. Consequently, feature level modification techniques are not directly usable to generate malware binary files. In one of the earlier works in modifying features of a malware using AI-based techniques, Grosse {\em et al.} ~\cite{grosse2016adversarial, grosse2017adversarial} generated adversarial malware instances using a gradient based attack where the input feature that causes a maximal positive gradient for a target class is perturbed. To ensure that the functionality of the binary is not altered due to the perturbation, they limited the number of features of the binary that could be perturbed as well as ensured that the perturbation resulted in modifying exactly one line of code in the original binary. Evaluation of the approach on the DREBIN dataset achieved $65-70\%$ misclassification rate for the perturbed binary with around $14$ features changed in each binary. The malware classifier was subsequently hardened using techniques including feature reduction, defensive distillation and adversarial training, albeit with small improvement in misclassification rates. Recently, in~\cite{anderson2018learning}, authors propose a technique using reinforcement learning to modify certain features in a malware binary so that it can evade a decision tree gradient boost-based malware classifier. They assume that features are functionality preserving (malware continues to act as a malware post-modification) and that the malware creator or adversary can observe the modification's effect (in other words, observe the classifier's output on modified malware). The malware modification process is implemented as an evasion game where the adversary makes multiple attempts to evade the classifier with the number of attempts limited by a certain threshold. Subsequently, they use adversarial training of the decision tree-based malware classifier with the modified samples. Results showed that the adversarial training improved accuracy of adversarial malware by $4\%$. The authors also point out some limitations of their approach, such as the adversarial training might end up learning to distinguish between modified by Library to Instrument Executable Formats (LIEF) versus not modified by LIEF instead of adversarial malware versus benign, and that lazy execution in Windows might be exploited to bypass their proposed adversarial modification techniques. Fang {\em et al.}~\cite{fang2021ac3mal} proposed another reinforcement learning technique called AC3Mal to determine a rule or policy that converts malware to adversarial by performing different actions on a binary~\cite{anderson2017evading} such as appending bytes, sections, or libraries, removing sections, and modifying the binary's signature certificate, to modify the features of the binary and make it evasive while preserving its functionality. 
Authors have also proposed generative adversarial networks (GANs)~\cite{hu2017generating}, enhanced with Monte Carlo Tree Search~\cite{zhang2020adversarial} to generate adversarial malware by modifying features of the binary.

{\bf Strategically Modifying Bytes in Malware Binary.} Byte-level modifications of a malware binary are effective to alter their behavior. However, arbitrarily modifying bytes could break the functionality of the malware and render them un-executable. Researchers have addressed the functionality preserving issue either implicitly by preventing modification to sections within the binary that could affect its functionality~\cite{luca2019explaining}, or, explicitly by re-running the modified binary inside a sandbox environment and discarding it if it does not have the same functionality as the original~\cite{jin21fumvar}. Boutaskis {\em et al.}~\cite{boutsikas2021evading} proposed a malware generation technique where modifications such as adding strings, bytes, sections, or functions, and, changing the file's timestamp and signature were performed intelligently on a binary malware using a Monte Carlo Tree Search (MCTS) algorithm. Their results showed that the MCTS-based technique was able to make only a single modification to a binary file in more than $50\%$ of the tested binary file samples, so that the modified binary file could successfully evade a surrogate classifier{\footnote{A surrogate classifier is a classifier that is trained by the adversary using a smaller subset of the dataset used to train the classifier inside an AV tool, because the adversary might not have access to the AV tool's classifier.}}. When tested with the actual AV classifier, the modified binary's evasiveness was reported to decrease to $~9\%$. Similar approaches of modifying the binary, albeit using genetic algorithms such as AIMED~\cite{castro2019aimed} and FUMVar~\cite{jin21fumvar} have also been proposed. In~\cite{kucuk2020deceiving}, authors propose three types of adversarial attacks that are again based on genetic algorithms for intelligently modifying a PE malware binary's opcodes, API calls and system calls. Their experiments show that the modified binaries can evade random forest based classifiers with an evasion rate between $75-91\%$ for the three types of attacks.

Another direction of generating malware leverages modifying certain bytes inside a binary or adding byte sequences to a malware so that the resulting modified malware can evade detection~\cite{labaca2019poster,chen2019adversarial,castro2019armed}. In one of the first works in this direction, Kreuk {\em at al.}~\cite{kreuk2018deceiving} proposed appending bytes to a binary and modifying the appended bytes using the gradient of the malware classifier's loss function until the modified binary was able to evade the classifier. Kolosnjaji {\em et al.} improved on this approach while minimizing the number of appended bytes that need to be modified to effect evasion. In ~\cite{suciu2019exploring}, authors proposed a gradient based technique for modifying padding or slack bytes inserted by a compiler between different sections of a binary. The Genetic Adversarial Machine learning Malware Attack (GAMMA)~\cite{demetrio2021functionality} uses a genetic algorithm to identify locations within a malware binary file to inject with portions from a benign binary while reducing the number of injections. In contrast to adding and modifying benign bytes to a malware, adding malicious byte sequences to a benign binary was proposed in~\cite{fleshman2018static}. In ~\cite{ebrahimi2020binary}, authors proposed a recurrent neural network -based framework called MalRNN that learns to append suitable bytes to malware binaries and make them evasive. Evaluation of MalRNN on around $6000$ malware samples showed that appending a little over $5$ KB bytes to most binaries achieved over $80\%$ evasion on a commercial AV tool and between $75-10\%$ evasion rate on a neural network based classifier called MalConv and its variants while outperforming other comparable byte append techniques. Demtetrio {\em et al.} determined that malware classifiers like MalConv mainly used the bytes in the DOS header of a binary to discriminate between malware and clean binaries. Based on this, they proposed a technique inspired from ~\cite{kolosnjaji2018adversarial} that modifies the closest header byte that would maximally increase the probability of evasion~\cite{luca2019explaining}. The same authors proposed another technique that makes a malware evasive by modifying or extending the DOS header along with shifting the content of the first section~\cite{demetrio2021adversarial}. More recently, Lucas {\em et al.} have described a technique of first disassembling the binary, followed by strategically replacing or displacing portions of bytes to generate evasive malware. Evaluation of their approach on malware classifiers MalConv and Avast~\cite{krvcal2018deep} showed that combined replacing and displacing was successful at making nearly $100\%$ of the tested malware evasive to the classifiers, with only $1$ modification making a little over $50\%$ of the malware evasive, and about $100$ modifications making $90\%$ malware evasive.

\subsection{Supervised Learning-based Malware Classification}

\begin{figure}
\begin{center}
\includegraphics[width=2.5in]{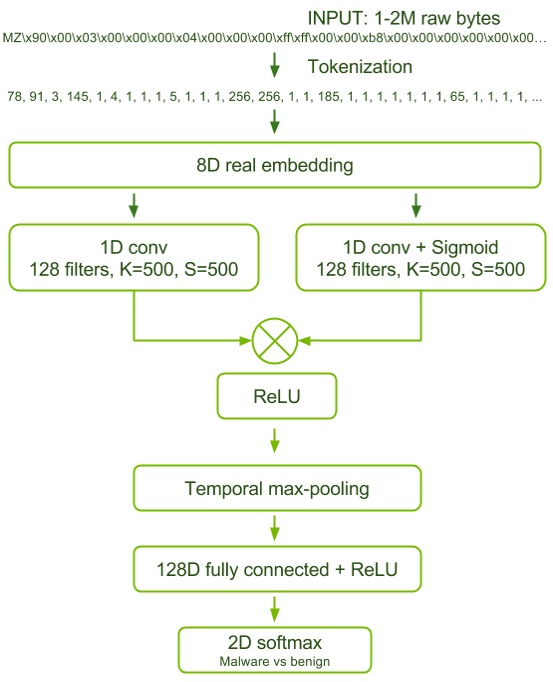}
\end{center}\caption{Deep neural network architecture used in MalConv~\cite{raff2018malware}.}
\label{fig_malconv}
\end{figure}
Initial approaches using ML-based malware classification~\cite{raman2012selecting,yan2013exploring} described methods to identify malware by carefully selecting and analyzing their content (e.g., byte strings, n-grams) and/or features, followed by supervised learning on the selected features. Several researchers proposed deep learning models that learn to classify malware based on features including numerical PE fields~\cite{saxe2015deep}, features extracted from dynamic analysis in a sandbox~\cite{david2015deepsign}, API calls~\cite{hardy2016dl4md} and opcodes~\cite{mclaughlin2017deep}. One of the first and widely used techniques to classify malware using the byte-level features from binary files uses a $4$-layer neural network called MalConv~\cite{raff2018malware} to train a classifier model using malware binary samples. A schematic of the deep neural network used in MalConv is shown in Figure~\ref{fig_malconv}. The trained classifier achieved an accuracy between $88-94\%$. Malconv was shown to be able to successfully identify new, previously unseen malware signatures. In contrast to reports in previous literature that the PE file header only contributed to a malware's signature, MalConv showed that different parts of the binary such as the .data, .text, .rdata sections, had a small but non-negligible contribution to the malware's detection by the classifier. A limitation with Malconv is that it only classifies the first $256$ bytes or $2$MB of a binary file, although this limitation was addressed recently~\cite{RaffFZAFM21}. In ~\cite{krvcal2018deep}, authors reported that extending MalConv's deep neural network architecture with convolution layers with strides of $4$ and $8$ right after the input embedding layer, and fully connected layers near the output, resulted in moderate performance improvements including about $1.5\%$ improvement in accuracy on a proprietary malware dataset. Coull and Gardner~\cite{coull2019activation} describe another deep neural network for byte-level malware classification along with investigating the effect of byte locations in the malware on the activations across different levels in the neural network.

\subsection{Malware Data Sets}
One of the major challenges with using AI for malware generation and detection, is the relative scarcity of reliable and publicly available malware data sets for training and evaluating the AI techniques. Table~\ref{table_malware_datasets} gives a comparison of the malware binary datasets in the public domain that are widely used for research. Another widely used malware dataset and detection service called VirusTotal is available publicly for academic use only; SOREL $20$M dataset mentions that most samples from VirusTotal are already included in their dataset. Several sources for free malware samples for researchers are also provided at~\cite{lennyzeltser2021}.

\begin{table}
\begin{tabular}{|c|c|c|}
\hline
{\bf Name} & {\bf Description} & {\bf Data Size}\\
\hline
DREBIN~\cite{arp2014drebin}		& Android malware	 	& $5560$ samples  from $179$ \\
(2010-2012)						&						&  malware families \\
\hline
EMBER $2018$ ~\cite{anderson2018ember} & Features from headers 	& Train set: $800$K ($300$K , \\
(2017-2018)						& of Win PE files 		& benign $300$K malware, $200$K \\
								&						& unlabeled; Test set: $200$K\\
\hline
SOREL $20$M~\cite{harang2020sorel}	& Binary malware Win & $20$M features, $10$M binary\\
(2020)								& PE files; disarmed & benign\\
\hline
vx-underground~\cite{vxug2021} 	& Android, linux, Win	 & $>1$M samples, new samples \\
								& PE files; not disarmed & added regularly\\
\hline								
Virus Share~\cite{virusshare2021} & Android, linux, Win  & New samples added\\
								& PE files; not disarmed & regularly\\
\hline
Malware Share~\cite{malshare2021} & Android, linux, Win  & New samples added \\
								& PE files; not disarmed & regularly\\							
\hline
\end{tabular}
\caption{Malware binary datasets available in public domain that have been used in adversarial malware generation research.}
\label{table_malware_datasets}
\end{table}

\section{Adversarial Malware Generation}
\label{sec_model}
\begin{figure}[tbh!]
\begin{center}
\includegraphics[width=4.0in]{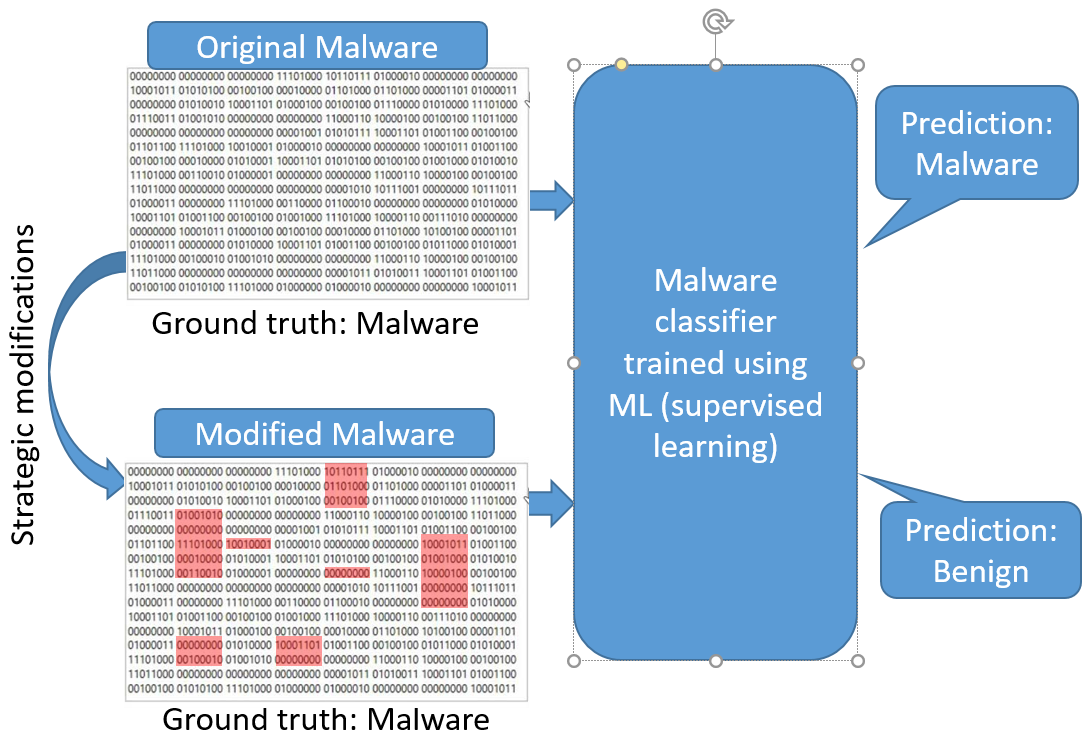}
\end{center}
\caption{Schematic of adversarial malware generation. The original malware (top left) is modified by adding, removing or modifying some bytes to create a modified malware such that the modified malware is able to evade classification.}
\label{fig_amg_schematic}
\end{figure}
The general concept of creating an adversarial or evasive version of a malware binary is to selectively modify, add or remove bytes in the original malware until the modified malware is able to avoid detection by a malware detector like an AV tool or a malware classifier{\footnote{We assume that the malware classifier has been pre-trained to classify malware with acceptable accuracy. Also, following literature on adversarial malware generation, we only consider evasion attacks on the malware classifier. Poisoning attacks could be effected using adversarial generation techniques by introducing adversarial malware into the training set used to train the malware classifier.}}. A schematic of this approach is shown in Figure~\ref{fig_amg_schematic}. A binary file consists of a sequence of bytes. Formally, let $B = \{b_1, b_2, ...b_{|N|}\}$ denote the byte sequence of a binary file $B$, where $|B|$ denotes the number of bytes of the binary. Let $f(B) = \{0, 1\}$ denote the functionality of the binary file where $0$ means a benign functionality and $1$ means a malicious functionality{\footnote{The functionality of $B$, $f(B)$, can be determined by running $B$ within its intended operating environment like Windows, Linux, Android, etc., or, in case $B$ is known to be a malware binary, within an appropriate sandbox emulating $B$'s intended operating environment.}}. For example, the functionality of a benign binary like {\tt wingrep} is to implement the {\tt grep} function on a Windows operating system, while the functionality of a malware binary like the {\em Code Red} computer worm is to cause a program buffer overflow and allow it to execute arbitrary code. For notation convenience, we write $f(B_1) = f(B_2)$ if two binary files, $B_1$ and $B_2$, have the same functionality, although $f()$ might be implemented differently for $B_1$ and $B_2$. Let ${\cal C}: B \rightarrow \{0, 1\}$ denote a classifier that takes a binary file as input and returns a binary output denoting whether the binary file is benign($0$) versus malware($1$).{\footnote{We assume that ${\cal C}$ is an oracle, that is $f(B) = C(B),\,\forall B$. In practice, we could determine both ${\cal C}(B)$ and $f(B)$ and retain $B$ only if $f(B) = C(B)$.}} For notation convenience, we denote a binary that is classified by ${\cal C}$ as malware by $B_{mal}$. Let $\phi: B \rightarrow B$ denote a function that takes a binary file $B = \{b_1, ..., b_{j_1},..., b_{j_2},..., b_{|B|}\}$  and outputs another binary file $B' = \{b_1, ..., b'_{j_1}, ..., b'_{j_2},..., b'_{|B'|}\}$, where bytes at indices $j_1, j_2, ...$ are modified between $B$ and $B'$. Note that $|B| <> |B'|$, as $\phi()$ could append or remove bytes in $B$ to create $B'$. Let $n(\phi)$ denote the number of byte modifications, additions and removals made by $\phi()$ to modify $B$ into $B'$. The adversarial malware generation problem can then be stated as:

\begin{eqnarray}
\min n(\phi) \label{amg_objf}\\
\mbox{s. t.}: \nonumber \\
B' = \phi(B), \label{amg_cons1}\\
{\cal C}(B) = 1, \label{amg_cons2}\\
{\cal C}(B') = 0, \label{amg_cons3}\\
f(B) = f(B') \label{amg_cons4}
\end{eqnarray}

In the above, the objective function in~\ref{amg_objf} is to reduce the cost of the adversary measured in terms of number of modifications to the original binary. Equations~\ref{amg_cons1}-~\ref{amg_cons3} specify that the modified binary should be able to evade the malware classifier, while Equation~\ref{amg_cons4} specifies the functionality preservation constraint.

Existing adversarial malware generation techniques have mainly developed algorithms that satisfy Constraints ~\ref{amg_cons1}-~\ref{amg_cons3}. Constraint~\ref{amg_cons4} (functionality preserving modification) is guaranteed either implicitly by preventing $\phi()$ from modifying sections of $B$ that could affect its functionality~\cite{luca2019explaining}, or, explicitly by re-running $B'$ inside a sandbox environment and discarding $B'$ if it does not have the same functionality as $B$~\cite{jin21fumvar}. Many existing techniques, however, do not attempt satisfy the objective function in Equation~\ref{amg_objf}; $n(\phi)$ is usually measured in terms of number of steps or iterations required by the algorithm implementing $\phi()$. We selected three techniques for adversarial malware generation from recent literature based on their reported performances along these two metrics, as described below:

\subsection{Padding Attack~\cite{kolosnjaji2018adversarial}}

\begin{figure}[thb!]
\begin{tabular}{cc}
\includegraphics[scale=0.8]{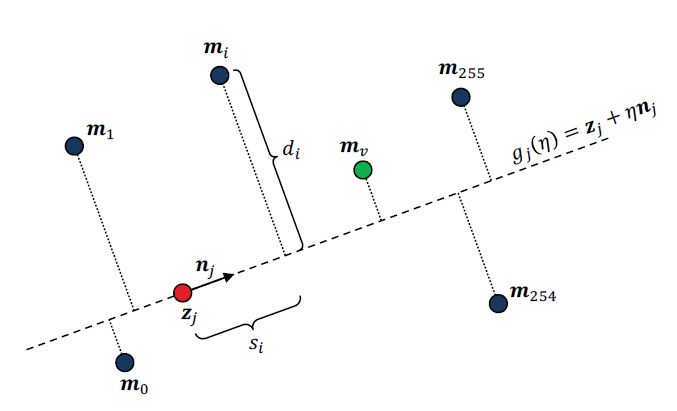}&
\includegraphics[scale=0.8]{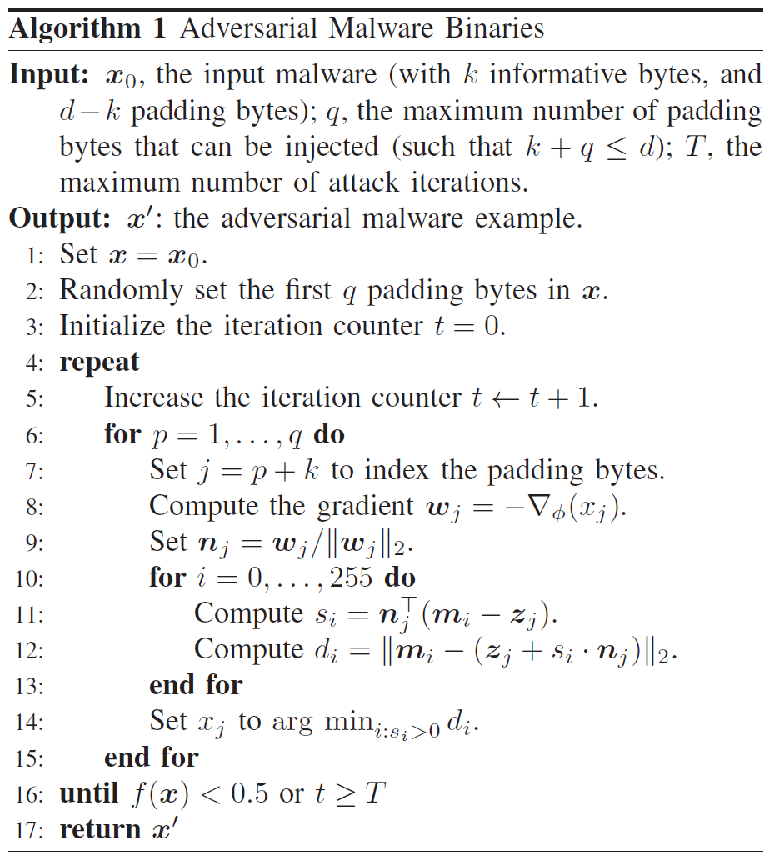}
\end{tabular}
\caption{Schematic of gradient-based padding byte modification (left) and algorithm (right) for modifying malware described in Kolosnjaji {\em et al.}~\cite{kolosnjaji2018adversarial}.}
\label{fig_kolosnjaji}
\end{figure}

The objective of the padding attack is to modify a malware binary by strategically appending a certain number of bytes, called padding bytes to it, so that the confidence of the classifier for the modified binary reduces to $<50\%$. Padding bytes are initialized randomly. Each padding byte is then substituted by the byte from the original malware that has the minimum distance (measured as L2-norm) from the padding byte, with distance measured in the direction of the negative gradient of the padding byte with respect to the classifier's model. The byte replacements are continued until the classifier confidence falls below $50\%$, or, a maximum number of replacement iterations is exceeded. The padding attack is a white-box attack as it requires knowledge of the malware classifier's model to calculate the gradient for the padding bytes' modifications. A graphical representation of the attack and pseudo-code algorithm are shown in Figure~\ref{fig_kolosnjaji}. The proposed approach was validated by the authors on the MalConv classifier~\cite{raff2018malware} with malware samples from VirusShare, Citadel, APT1. Results showed that gradient based modification on appended bytes were more effective than random modifications as fewer, targeted bytes need to be changed to make a malware evasive. Also, modifying first few bytes in a binary was found to be more effective in making the malware evasive than modifying later bytes.

\subsection{Partial DOS Header Manipulation~\cite{luca2019explaining}}
\begin{figure}[thb!]
\begin{tabular}{cc}
\includegraphics[scale=1.0]{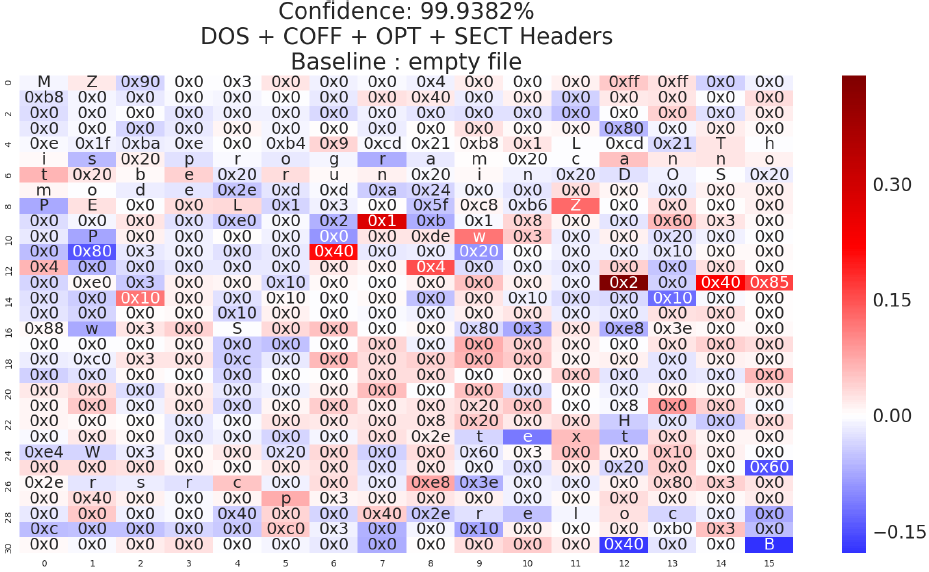}&
\includegraphics[scale=0.8]{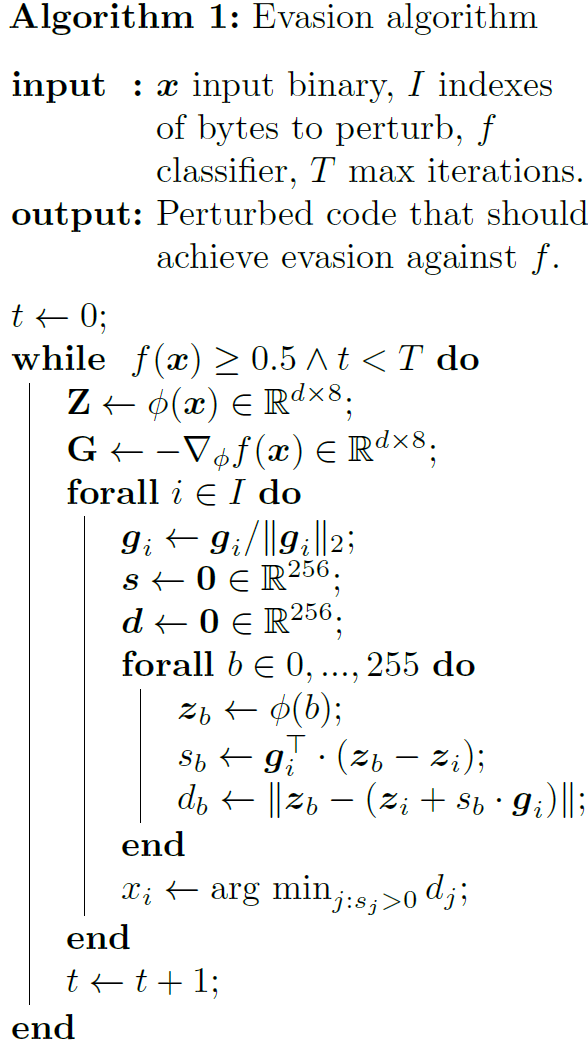}\\
{\small{(a)}}& {\small{(b)}}
\end{tabular}
\caption{(a) Attribution of different bytes in a binary file's header towards classifying the binary file as malware using MalConv. (b) Algorithm for modifying header bytes in the original malware with high malware attribution. The image and algorithm are from~\cite{luca2019explaining}.}
\label{fig_dos_demetrio}
\end{figure}

The DOS header manipulation attack works by identifying the attributions of different features or bytes in a binary's header towards classifying it as malware versus benign using MalConv.{\footnote{This attack was mainly developed to explain the inner workings of the MalConv classifier and understand its vulnerabilities.}} The attribution of input $\mathbf{x}$ with respect to a baseline $\mathbf{x'}$ along feature $i$ is given in terms of a metric called integrated gradient~\cite{sundararajan2017axiomatic}, that is defined as:
\begin{equation}
IG_i(\mathbf{x})  = (x_i - x'_i)\int_0^1 \frac{\partial f(\mathbf{x'} + \alpha (\mathbf{x'}-\mathbf{x'}))}{\partial x_i} d\alpha,
\end{equation}

\noindent where $f()$ is the classifier's model and $\alpha$ is a constant. An example of the attribution of different bytes in a binary file's header is shown in Figure~\ref{fig_dos_demetrio}(a). The bytes that are identified to have the highest attribution towards a binary being classified as malware are then modified using a gradient-based approach similar to the technique used to modify padding bytes in the aforementioned padding attack~\cite{kolosnjaji2018adversarial}. The DOS header attack is also a white-box model as it requires knowledge of the classifier's model to determine the header bytes to modify as well as to calculate the gradients for the modification. The attack was validated by the authors on $60$ malware samples obtained from Websites {\em The Zoo} and {\em Das Malwerk} and their results showed that a little over $8000$ header byte modifications on a binary (~$150$ iterations of the outer loop in the algorithm in Figure~\ref{fig_dos_demetrio}(b), with each iteration modifying $58$ header bytes) were able to reduce MalConv's confidence of classifying the binary as malware below $50\%$ for $52$ of the $60$ evaluated malware samples.

\subsection{Genetic Adversarial Machine learning Malware Attack (GAMMA)~\cite{demetrio2021functionality}}

\begin{wrapfigure}{r}{0.5\textwidth}
\begin{center}
\includegraphics[scale=1.0]{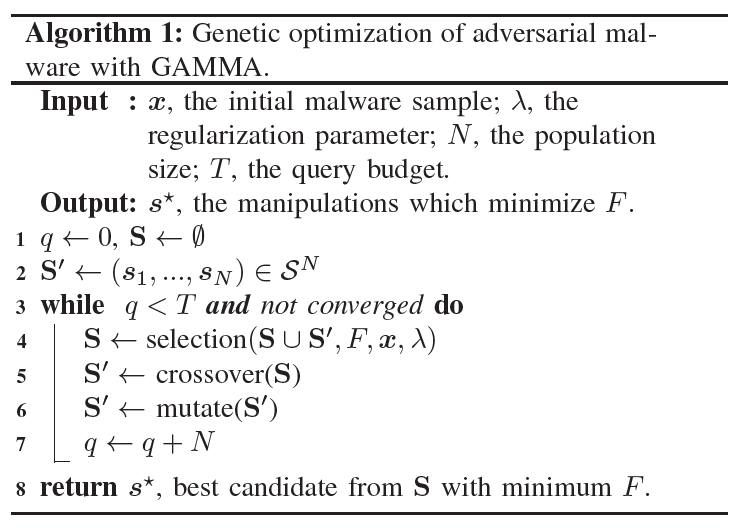}
\end{center}
\caption{Genetic Algorithm Malware Attack (GAMMA) algorithm.}
\label{fig_gamma_algo}
\end{wrapfigure}

\begin{figure}[htb!]
\begin{center}
\includegraphics[scale=1.0]{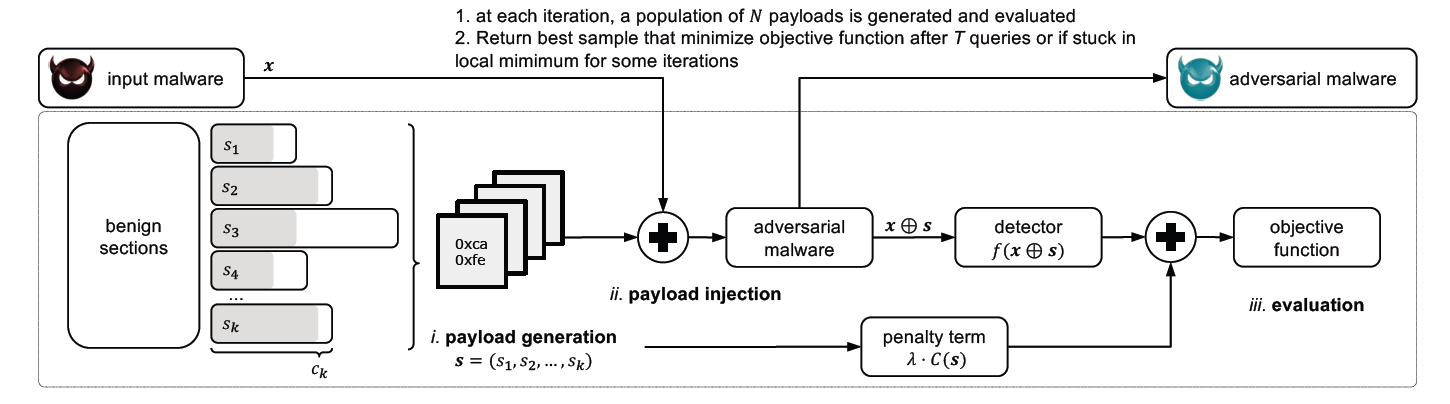}
\end{center}
\caption{Schematic for the Genetic Algorithm Malware Attack (GAMMA) technique.}
\label{fig_gamma_schematic}
\end{figure}

GAMMA was developed as an attack that is less detectable by humans than the DOS header attack. The main idea in GAMMA is to determine the fraction of bytes to modify within each section of the malware binary file using a genetic algorithm until the modified binary is able to evade detection, while reducing the number of modifications. GAMMA considers two types of modifications - adding benign bytes into each section of the binary via padding attack and adding new sections into the binary via section injection attack. An upper threshold or budget of $510$ modifications is allowed on the binary being modified. GAMMA is mentioned as black box attack as the attacker does not need access to the classifier's model, although it does need to query the classifier repeatedly until either the modified malware is able to evade the classifier or the modification budget is reached. A schematic of the GAMMA technique is shown in Figure~\ref{fig_gamma_schematic} and the corresponding algorithm is shown in Figure~\ref{fig_gamma_algo}. Malware modified by GAMMA was evaluated with a Gradient-based Decision Tree classifier and Malconv by the authors. Both classifiers were trained with $15000$ benign and malware samples each obtained from Virustotal and evaluated with $500$ malware samples. Results showed that modifying between $500$ to $1000$ bytes of the original malware was successful in increasing the evasion rate by an average of $40\%$ for both classifiers and attacks. 

\section{Experimental Results}
\label{sec_expt}
{\bf Data sets used for Experiments.} For evaluating the adversarial malware generation techniques discussed above we used malware samples obtained from vx-underground~\cite{vxug2021}. We used Block.$0115$ that contained $38482$ malware samples. We used the first $500$ files, analyzed the magic bytes of the file using the UNIX {\tt file} command to ascertain the file type and retained only Windows files. This gave us a data set of $306$ Windows PE malware samples.  It is possible for a malware sample to be "hiding" it's true nature by falsifying its magic bytes, but we think this is unlikely as it would impede the malware's functionality. {\footnote{For the SOREL-$20$M dataset~\cite{harang2020sorel} that contains disarmed malware binaries, we found that most of the malware samples being mis-classified as benign by MalConv. Among the remaining binaries most were being made evasive in only one iteration by the DOS header attack. We did not use the SOREL dataset for our experiments as we felt that modifying the header to disarm  the samples might have effected their mis-classification and relative ease in converting to being evasive.}} We also created $13000$ benign Windows PE binaries consisting of .exe and .dll files extracted from different Windows software. Portions of these binaries were used to inject the malware while using GAMMA. 

\noindent {\bf Software and Python Scripts.} We used the {\tt secml\_malware} codebase version 0.2.4 provided by Demetrio and Biggio~\cite{demetrio2021secmlmalware} and made the following modifications for the purpose of our project:
\begin{enumerate}
\item A script named \verb|attack.py| that allows programmers to pass the attack type and the maximum number of samples to be tested as parameters. The generated adversarial binaries are saved at the file path provided in the \verb|save_dir| argument to the script.
\item A helper script caleed \verb|testing.py| which can batch run all of the tests, and save the output to a file.
\item A script named \verb|csvgen.py| that generates comma separate text (csv) files from the output generated by \verb|testing.py|. 
\item A Jupyter notebook in the codebase, named \verb|attack_evasion_examples.ipynb|. This notebook contains documentation and example usage of how to use the script created, and how to build things using the script. 
\item A \verb|req.install| file that contains the software dependencies across the different software libraries used in the codebase for easier reproducibility.
\end{enumerate} 

The codebase was run using Python $3.9.6$ and Tensorflow $1.15$. We used MalConv as the malware classifier with a pre-trained model that was trained using the EMBER dataset\cite{anderson2018ember} and available from~\cite{embergithub2018}. The confidence threshold for a binary to be classified as malware was set to $0.5$, following the findings reported in~\cite{demetrio2021functionality}. The experiments were performed on a desktop machine with a $i7-8700k$ CPU, Nvidia RTX3060TI GPU, $16$GB RAM and running Manjaro Linux.

\noindent {\bf Evaluation Metrics.} We used the following three measures to evaluate performance of the compared techniques: 
\begin{itemize}
    \item Evasion rate: The evasion rate is the fraction of malware binaries that were able to evade the MalConv classifier post-modification. Mathematically, it can be denoted as $\rho_{eva} = \frac{|\mathbf{B'}|}{|\mathbf{B}|}$, where $\mathbf{B}$ is a set of binary malware, $\mathbf{B}' = \{\phi(B), B \in \mathbf{B} \wedge {\cal C}(\phi(B))=0\}$, and, $\phi()$ and ${\cal C}()$ are the adversarial malware generation function and malware classifier respectively. Most existing adversarial malware generation techniques use evasion rate as the main performance measure to evaluate the effectiveness of their approach.
    \item Time taken (in seconds) to make a binary evasive
    \item Number of perturbations to make a binary evasive
\end{itemize}

\begin{figure}
\begin{tabular}{cc}
\includegraphics[scale=.5]{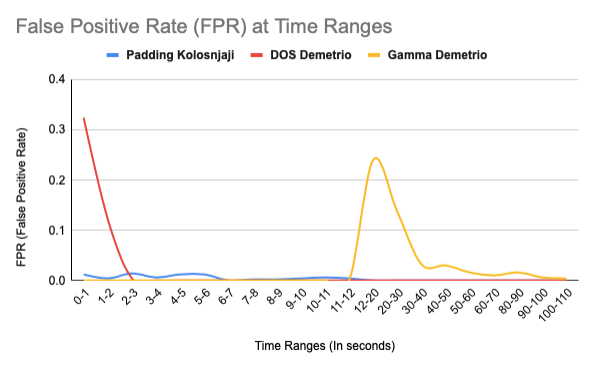} &
\includegraphics[scale=.5]{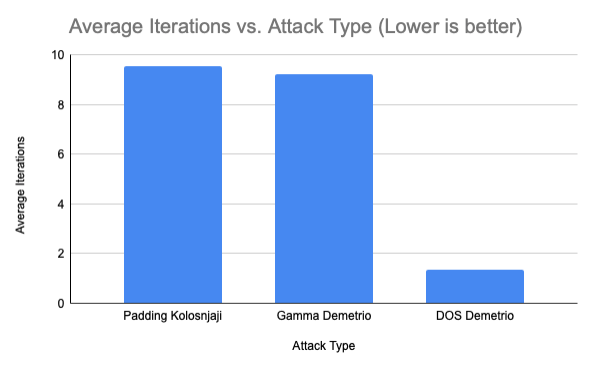}\\
{\small{(a)}} &{\small{(b)}}\\
\includegraphics[scale=.5]{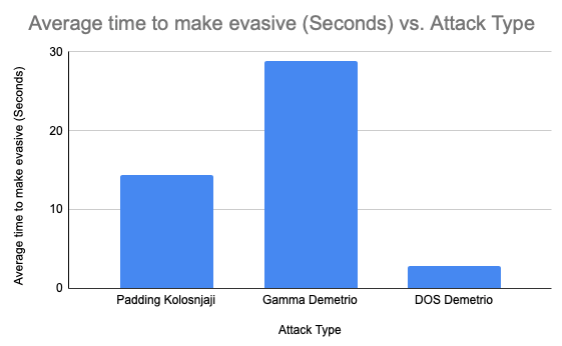} &
\includegraphics[scale=.5]{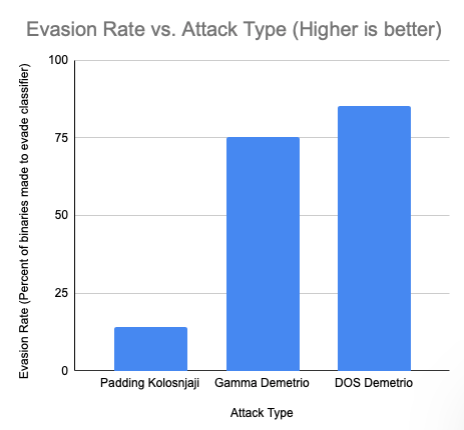}\\
{\small{(c)}} &{\small{(d)}}
\end{tabular}
\caption{Results of the implemented techniques on the vx-underground malware data. (a) False positive rate, (b) Number of iterations, (c) Processing time, (d) Evasion Rate for the three evaluated techniques.}
\label{fig_amg_results}
\end{figure}

\noindent {\bf Results.} The results from our experiments are shown in Figure~\ref{fig_amg_results}(a) - (d). We can see that Partial DOS Header Attack (DOS) was by far the most effective with the higher evasion rate, lowest average time and lowest number of perturbation, among all three attacks. The second best results were for GAMMA, which was able to preform nearly as well as DOS Header Attack in terms of the evasion rate but took considerably longer time to make samples evasive. The padding attack performed the least effectively among the three evaluated attacks.

\section{Lessons Learned and Future Work} 
\label{sec_lessons}
In this section, we highlight some interesting issues that we learned from this research and identify future directions based on those issues that would be worthwhile for researchers to investgate.

{\bf Correlation Between Binary File Features and Attacks.} An interesting feature during testing was that some attack methods were effective on some files where others were not. For example, while the DOS header attack was able to make some samples adversarial the GAMMA attack would fail on the same sample, and vice-versa. Exploring what features or signature in the binary makes them suitable for evasion using one type of attack over another appears; effectively answering the question. "Why do some attacks work better on some binaries than others?" is an interesting direction worthy of further investigation. A continuation of this research question: "What attacks works best on X malware sample?" towards building an ensemble classifier that would be robust to different adversarial malware binaries.

{\bf Binary Functionality from Static Analysis.} Another interesting question we ran into, "Can we statically determine if two binaries are semantically the same?", while in appearance they look different. We suspect this alone is a separate field of research, but if we were able to create a way to reliably test this, making sure the binary modifications we do to adversarial samples do not actually break the binary would become trivial. Due to the fact that machines execute these binaries in a very specific and non-random pattern, on first glance it seems to be possible to develop a software to do this.

{\bf Variations from Results in Literature.} One of the very interesting behaviors that came up during our experiments was that any sample that was classified with $100\%$ confidence by the MalConv classifier was not able to be modified by any of the evaluated attacks into an evasive malware. This requires further investigation. Another interesting behavior observed Is a divergence from the original paper, Demetrio et al shows a graph depicting number of iterations preformed vs evasion rate, this graph showed evasion rate falling drastically around 150 iterations. In our own findings however we show most binaries being made to bypass the classifier within the first 3 iterations, most of them being made adversarial in 1 or 2. As of yet we have no explanation for the difference in behavior we record compared to the results of the author. 

Currently, we are reproducing the results of some recent adversarial malware generation technique including Malware-makeover~\cite{lucas21makeover}, AC3Mal~\cite{fang2021ac3mal} and MalRNN~\cite{ebrahimi2020binary} with the dataset and codebases used in this report.

Overall, in this research we implemented three adversarial malware generation techniques from recent literature and evaluated their effectiveness on the same dataset and while using the same malware classifier. Our results showed that the DOS Header attack was the most effective technique. While recent research in adversarial AI has focused on image and text domains, commensurate techniques for binary files for generating adversarial malware have been less prevalent. As noted in~\cite{raff2018malware}, the difficult of acquiring reliably labeled binary malware sample and the uniqueness of the features and behavior of binary data as compared to image and text data, have been challenges for adversarial malware research. This research is our first step in addressing some of these challenges and we envisage that the results and lessons from our research will lead to addressing more challenging issues in this important topic.

\section{Contributions}
Dasgupta was responsible for supervising the research and writing the material in Sections~\ref{sec_intro}-~\ref{sec_model} of this report. Osman was responsible for implementing the software and experiments reported here, as well as for the material in Sections~\ref{sec_expt}-~\ref{sec_lessons}.

\section{Acknowledgements}
The authors would like to thank the U. S. Office of Naval Research for supporting this research through a NRL Base Funding grant to Dr. Dasgupta. The authors would also like to thank Keane Lucas, Ph.D. student at Carnegie Mellon University and first author of~\cite{lucas21makeover}, for providing valuable insights, comments, and pointers to the software implementation and data sets of~\cite{lucas21makeover}.

\bibliographystyle{abbrv}

\begin{thebibliography}{10}

\bibitem{anderson2018learning}
H.~S. Anderson, A.~Kharkar, B.~Filar, D.~Evans, and P.~Roth.
\newblock Learning to evade static pe machine learning malware models via
  reinforcement learning.
\newblock {\em arXiv preprint arXiv:1801.08917}, 2018.

\bibitem{anderson2017evading}
H.~S. Anderson, A.~Kharkar, B.~Filar, and P.~Roth.
\newblock Evading machine learning malware detection.
\newblock {\em Black Hat}, pages 1--6, 2017.

\bibitem{anderson2018ember}
H.~S. Anderson and P.~Roth.
\newblock Ember: An open dataset for training static pe malware machine
  learning models. arxiv e-prints (april 2018).
\newblock {\em arXiv preprint arXiv:1804.04637}, 2018.

\bibitem{malshare2021}
Anonymous.
\newblock Malware share.
\newblock https://www.malshare.com, 2021.

\bibitem{virusshare2021}
Anonymous.
\newblock Virus share.
\newblock https://www.virusshare.com, 2021.

\bibitem{vxug2021}
Anonymous.
\newblock Vx-underground.
\newblock https://www.vx-underground.org, 2021.

\bibitem{arp2014drebin}
D.~Arp, M.~Spreitzenbarth, M.~Hubner, H.~Gascon, K.~Rieck, and C.~Siemens.
\newblock Drebin: Effective and explainable detection of android malware in
  your pocket.
\newblock In {\em Ndss}, volume~14, pages 23--26, 2014.

\bibitem{boutsikas2021evading}
J.~Boutsikas, M.~E. Eren, C.~Varga, E.~Raff, C.~Matuszek, and C.~Nicholas.
\newblock Evading malware classifiers via monte carlo mutant feature discovery.
\newblock {\em arXiv preprint arXiv:2106.07860}, 2021.

\bibitem{castro2019aimed}
R.~L. Castro, C.~Schmitt, and G.~Dreo.
\newblock Aimed: Evolving malware with genetic programming to evade detection.
\newblock In {\em 2019 18th IEEE International Conference On Trust, Security
  And Privacy In Computing And Communications/13th IEEE International
  Conference On Big Data Science And Engineering (TrustCom/BigDataSE)}, pages
  240--247, 2019.

\bibitem{castro2019armed}
R.~L. Castro, C.~Schmitt, and G.~D. Rodosek.
\newblock Armed: How automatic malware modifications can evade static
  detection?
\newblock In {\em 2019 5th International Conference on Information Management
  (ICIM)}, pages 20--27. IEEE, 2019.

\bibitem{chen2019adversarial}
B.~Chen, Z.~Ren, C.~Yu, I.~Hussain, and J.~Liu.
\newblock Adversarial examples for cnn-based malware detectors.
\newblock {\em IEEE Access}, 7:54360--54371, 2019.

\bibitem{coull2019activation}
S.~E. Coull and C.~Gardner.
\newblock Activation analysis of a byte-based deep neural network for malware
  classification.
\newblock In {\em 2019 IEEE Security and Privacy Workshops (SPW)}, pages
  21--27. IEEE, 2019.

\bibitem{david2015deepsign}
O.~E. David and N.~S. Netanyahu.
\newblock Deepsign: Deep learning for automatic malware signature generation
  and classification.
\newblock In {\em 2015 International Joint Conference on Neural Networks
  (IJCNN)}, pages 1--8. IEEE, 2015.

\bibitem{demetrio2021secmlmalware}
L.~Demetrio and B.~Biggio.
\newblock secml-malware: Pentesting windows malware classifiers with
  adversarial exemples in python, 2021.

\bibitem{demetrio2021functionality}
L.~Demetrio, B.~Biggio, G.~Lagorio, F.~Roli, and A.~Armando.
\newblock Functionality-preserving black-box optimization of adversarial
  windows malware.
\newblock {\em IEEE Transactions on Information Forensics and Security},
  16:3469--3478, 2021.

\bibitem{demetrio2021adversarial}
L.~Demetrio, S.~E. Coull, B.~Biggio, G.~Lagorio, A.~Armando, and F.~Roli.
\newblock Adversarial exemples: A survey and experimental evaluation of
  practical attacks on machine learning for windows malware detection.
\newblock {\em ACM Transactions on Privacy and Security (TOPS)}, 24(4):1--31,
  2021.

\bibitem{ebrahimi2020binary}
M.~Ebrahimi, N.~Zhang, J.~Hu, M.~T. Raza, and H.~Chen.
\newblock Binary black-box evasion attacks against deep learning-based static
  malware detectors with adversarial byte-level language model.
\newblock 2020.

\bibitem{fang2021ac3mal}
Z.~Fang, J.~Wang, J.~Geng, Y.~Zhou, and X.~Kan.
\newblock A3cmal: Generating adversarial samples to force targeted
  misclassification by reinforcement learning.
\newblock {\em Applied Soft Computing}, 109:107505, 2021.

\bibitem{fleshman2018static}
W.~Fleshman, E.~Raff, R.~Zak, M.~McLean, and C.~Nicholas.
\newblock Static malware detection \& subterfuge: Quantifying the robustness of
  machine learning and current anti-virus.
\newblock In {\em 2018 13th International Conference on Malicious and Unwanted
  Software (MALWARE)}, pages 1--10. IEEE, 2018.

\bibitem{grosse2016adversarial}
K.~Grosse, N.~Papernot, P.~Manoharan, M.~Backes, and P.~McDaniel.
\newblock Adversarial perturbations against deep neural networks for malware
  classification.
\newblock {\em arXiv preprint arXiv:1606.04435}, 2016.

\bibitem{grosse2017adversarial}
K.~Grosse, N.~Papernot, P.~Manoharan, M.~Backes, and P.~McDaniel.
\newblock Adversarial examples for malware detection.
\newblock In {\em European symposium on research in computer security}, pages
  62--79. Springer, 2017.

\bibitem{harang2020sorel}
R.~Harang and E.~M. Rudd.
\newblock Sorel-20m: A large scale benchmark dataset for malicious pe
  detection.
\newblock {\em arXiv preprint arXiv:2012.07634}, 2020.

\bibitem{hardy2016dl4md}
W.~Hardy, L.~Chen, S.~Hou, Y.~Ye, and X.~Li.
\newblock Dl4md: A deep learning framework for intelligent malware detection.
\newblock In {\em Proceedings of the International Conference on Data Science
  (ICDATA)}, page~61. The Steering Committee of The World Congress in Computer
  Science, Computer~…, 2016.

\bibitem{hu2017generating}
W.~Hu and Y.~Tan.
\newblock Generating adversarial malware examples for black-box attacks based
  on gan.
\newblock {\em arXiv preprint arXiv:1702.05983}, 2017.

\bibitem{embergithub2018}
A.~Hyrum and P.~Roth.
\newblock Ember github.
\newblock https://github.com/elastic/ember, 2018.

\bibitem{jin21fumvar}
B.~Jin, J.~Choi, H.~Kim, and J.~B. Hong.
\newblock Fumvar: A practical framework for generating <u
  class="uu">f</u>ully-working and <u class="uu">u</u>nseen <u
  class="uu">m</u>alware <u class="uu">var</u>iants.
\newblock In {\em Proceedings of the 36th Annual ACM Symposium on Applied
  Computing}, SAC '21, page 1656–1663, New York, NY, USA, 2021. Association
  for Computing Machinery.

\bibitem{kolosnjaji2018adversarial}
B.~Kolosnjaji, A.~Demontis, B.~Biggio, D.~Maiorca, G.~Giacinto, C.~Eckert, and
  F.~Roli.
\newblock Adversarial malware binaries: Evading deep learning for malware
  detection in executables.
\newblock In {\em 2018 26th European signal processing conference (EUSIPCO)},
  pages 533--537. IEEE, 2018.

\bibitem{krvcal2018deep}
M.~Kr{\v{c}}{\'a}l, O.~{\v{S}}vec, M.~B{\'a}lek, and O.~Ja{\v{s}}ek.
\newblock Deep convolutional malware classifiers can learn from raw executables
  and labels only.
\newblock 2018.

\bibitem{kreuk2018deceiving}
F.~Kreuk, A.~Barak, S.~Aviv-Reuven, M.~Baruch, B.~Pinkas, and J.~Keshet.
\newblock Deceiving end-to-end deep learning malware detectors using
  adversarial examples.
\newblock {\em arXiv preprint arXiv:1802.04528}, 2018.

\bibitem{kucuk2020deceiving}
Y.~Kucuk and G.~Yan.
\newblock Deceiving portable executable malware classifiers into targeted
  misclassification with practical adversarial examples.
\newblock In {\em Proceedings of the Tenth ACM Conference on Data and
  Application Security and Privacy}, pages 341--352, 2020.

\bibitem{malwarebytes2020}
A.~Kujawa, N.~Collier, P.~Arntz, J.~Umawing, W.~Zamora, T.~Reed, J.~Segura, and
  C.~Boyd.
\newblock 2020 state of malware report.
\newblock
  https://www.malwarebytes.com/resources/2020-state-of-malware-report-pdf,
  2020.

\bibitem{labaca2019poster}
R.~Labaca-Castro, B.~Biggio, and G.~Dreo~Rodosek.
\newblock Poster: Attacking malware classifiers by crafting gradient-attacks
  that preserve functionality.
\newblock In {\em Proceedings of the 2019 ACM SIGSAC Conference on Computer and
  Communications Security}, pages 2565--2567, 2019.

\bibitem{luca2019explaining}
D.~Luca, B.~Biggio, L.~Giovanni, F.~Roli, and A.~Alessandro.
\newblock Explaining vulnerabilities of deep learning to adversarial malware
  binaries.
\newblock In {\em ITASEC19}, volume 2315, 2019.

\bibitem{lucas21makeover}
K.~Lucas, M.~Sharif, L.~Bauer, M.~K. Reiter, and S.~Shintre.
\newblock Malware makeover: Breaking ml-based static analysis by modifying
  executable bytes.
\newblock In {\em Proceedings of the 2021 ACM Asia Conference on Computer and
  Communications Security}, ASIA CCS '21, page 744–758, New York, NY, USA,
  2021. Association for Computing Machinery.

\bibitem{mclaughlin2017deep}
N.~McLaughlin, J.~Martinez~del Rincon, B.~Kang, S.~Yerima, P.~Miller, S.~Sezer,
  Y.~Safaei, E.~Trickel, Z.~Zhao, A.~Doup{\'e}, et~al.
\newblock Deep android malware detection.
\newblock In {\em Proceedings of the seventh ACM on conference on data and
  application security and privacy}, pages 301--308, 2017.

\bibitem{raff2018malware}
E.~Raff, J.~Barker, J.~Sylvester, R.~Brandon, B.~Catanzaro, and C.~K. Nicholas.
\newblock Malware detection by eating a whole exe.
\newblock In {\em Workshops at the Thirty-Second AAAI Conference on Artificial
  Intelligence}, 2018.

\bibitem{RaffFZAFM21}
E.~Raff, W.~Fleshman, R.~Zak, H.~S. Anderson, B.~Filar, and M.~McLean.
\newblock Classifying sequences of extreme length with constant memory applied
  to malware detection.
\newblock In {\em Thirty-Fifth {AAAI} Conference on Artificial Intelligence,
  {AAAI} 2021, Thirty-Third Conference on Innovative Applications of Artificial
  Intelligence, {IAAI} 2021, The Eleventh Symposium on Educational Advances in
  Artificial Intelligence, {EAAI} 2021, Virtual Event, February 2-9, 2021},
  pages 9386--9394. {AAAI} Press, 2021.

\bibitem{raman2012selecting}
K.~Raman et~al.
\newblock Selecting features to classify malware.
\newblock {\em InfoSec Southwest}, 2012:1--5, 2012.

\bibitem{saxe2015deep}
J.~Saxe and K.~Berlin.
\newblock Deep neural network based malware detection using two dimensional
  binary program features.
\newblock In {\em 2015 10th International Conference on Malicious and Unwanted
  Software (MALWARE)}, pages 11--20. IEEE, 2015.

\bibitem{suciu2019exploring}
O.~Suciu, S.~E. Coull, and J.~Johns.
\newblock Exploring adversarial examples in malware detection.
\newblock In {\em 2019 IEEE Security and Privacy Workshops (SPW)}, pages 8--14.
  IEEE, 2019.

\bibitem{sundararajan2017axiomatic}
M.~Sundararajan, A.~Taly, and Q.~Yan.
\newblock Axiomatic attribution for deep networks.
\newblock In {\em International Conference on Machine Learning}, pages
  3319--3328. PMLR, 2017.

\bibitem{yan2013exploring}
G.~Yan, N.~Brown, and D.~Kong.
\newblock Exploring discriminatory features for automated malware
  classification.
\newblock In {\em International Conference on Detection of Intrusions and
  Malware, and Vulnerability Assessment}, pages 41--61. Springer, 2013.

\bibitem{lennyzeltser2021}
L.~Zeltser.
\newblock Free malware sample sources for researchers.
\newblock https://zeltser.com/malware-sample-sources/, 2021.

\bibitem{zhang2020adversarial}
X.~Zhang, Y.~Zhou, S.~Pei, J.~Zhuge, and J.~Chen.
\newblock Adversarial examples detection for xss attacks based on generative
  adversarial networks.
\newblock {\em IEEE Access}, 8:10989--10996, 2020.

\end{thebibliography}

\end{document}